\documentclass[aip,jcp,twocolumn,10pt]{revtex4-1}
\usepackage{graphicx}%
\usepackage{dcolumn}
\usepackage{amsmath}   
\usepackage[usenames,dvipsnames]{pstricks}
\usepackage{epsfig}
\usepackage{pst-grad} 
\usepackage{pst-plot} 

\usepackage{graphicx} 
\usepackage{bm}
\usepackage{longtable}

\begin{document}

\title{Electric field inside a ``Rossky cavity'' in uniformly
  polarized water } 
\author{Daniel R.\ Martin}
\author{Allan D.\ Friesen}
\author{Dmitry V.\ Matyushov}
\email{dmitrym@asu.edu}
\affiliation{Center for Biological Physics, Arizona State University,
  PO Box 871604, Tempe, AZ 85287-1604 } 

\date{\today}

\begin{abstract}
  Electric field produced inside a solute by a uniformly polarized
  liquid is strongly affected by dipolar polarization of the liquid at
  the interface. We show, by numerical simulations, that the electric
  ``cavity'' field inside a hydrated non-polar solute does not follow
  the predictions of standard Maxwell's electrostatics of
  dielectrics. Instead, the field inside the solute tends, with
  increasing solute size, to the limit predicted by the Lorentz
  virtual cavity.  The standard paradigm fails because of its reliance
  on the surface charge density at the dielectric interface determined
  by the boundary conditions of the Maxwell dielectric. The interface
  of a polar liquid instead carries a preferential in-plane
  orientation of the surface dipoles thus producing virtually no
  surface charge.  The resulting boundary conditions for electrostatic
  problems differ from the traditional recipes, affecting the
  microscopic and macroscopic fields based on them. We show that
  relatively small differences in cavity fields propagate into
  significant differences in the dielectric constant of an ideal
  mixture. The slope of the dielectric increment of the mixture versus
  the solute concentration depends strongly on which polarization
  scenario at the interface is realized. A much steeper slope found in the
  case of Lorentz interfacial polarization also implies a higher free energy
  penalty for polarizing such mixtures.
\end{abstract}

\keywords{Cavity field, surface polarization, hydration, polar response. }
\maketitle

\section{Introduction}
\label{sec:1}
When an interface is created in a dielectric, the surface dipoles
change their preferential orientations relative to the dipoles in the
bulk. The response of the dielectric to a weak external field is then a
composite result of the response of these surface dipoles and the
bulk dipoles. The question of whether the dielectric response of a
material is sensitive to its surface structure is decided by the
relative weights of these two contributions.

\begin{figure}
  \centering
  \includegraphics*[width=8cm]{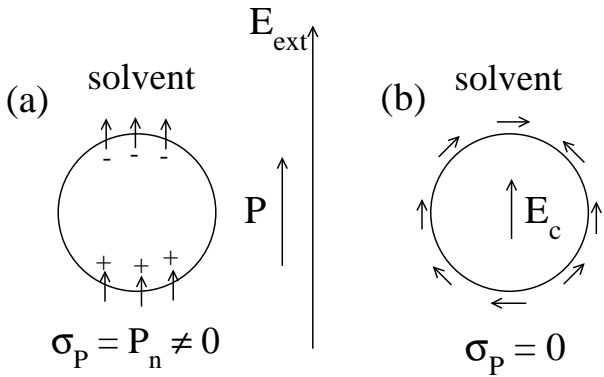}
  \caption{Cartoon of the charge distribution $\sigma_P$ at the
    surface of a spherical cavity carved from a liquid
    dielectric. Panel (a) represents the picture of standard Maxwell's
    electrostatics of dielectrics, when the normal projection of the
    dipolar polarization $\mathbf{P}$ along the external field
    $\mathbf{E}_{\text{ext}}$ creates the positive and negative lobes
    of the surface charge distribution. Those result from the
    mathematical surface cutting through polarized dipoles of the
    liquid shown by arrows. The overall electric field inside the
    cavity is then reduced from the external field by the opposing
    field of the surface charges [Eq.\ \eqref{eq:1}]. Panel (b) shows
    the scenario suggested by the orientational dipolar order at the
    surface of a ``Rossky cavity" in a polar
    liquid.\cite{Lee:84,Lee:86,Valleau:87,Sokhan:97,Bratko:09,Rossky:10,Romero-Vargas-Castrillon:2011mz}
    In-plane orientations of the surface dipoles, unaltered by a weak
    external field, do not produce surface charge. The result is zero
    surface charge density $\sigma_P$ and the field at the cavity
    center following the Lorentz equation [Eq.\ \eqref{eq:2}]. 
    Both sets of lines have been produced with $\epsilon = 72.2$ for
    SPC/E water at 300 K. }
  \label{fig:1}
\end{figure}

The Maxwell electrostatics of dielectrics neglects the structure of
the interface and replaces the interfacial region of a finite
microscopic dimension with an infinitesimally thin mathematical
surface.  This mathematical surface cuts through the dipoles of the
medium creating a surface charge with the charge density $\sigma_P$
(Fig.\ \ref{fig:1}a). It is given by the projection of the dipolar
polarization at the interface $\mathbf{P}(\mathbf{r}_S)$ on the
outward normal $\mathbf{\hat n}$ to the surface bounding the
dielectric,\cite{Landau8} $\sigma_P(\mathbf{r}_S)=P_n(\mathbf{r}_S)$,
$P_n(\mathbf{r}_S)=\mathbf{\hat n}\cdot \mathbf{P}(\mathbf{r}_S)$.

The surface charge density produces the electric field of its own,
which polarizes the dielectric to form an inhomogeneous polarization
$\mathbf{P}(\mathbf{r})$ near the interface.  It decays to the uniform
polarization field $\mathbf{P}$, associated with a uniform external
field $\mathbf{E}_{\text{ext}}$, far from the interface. Since the
spatial extent of the interface is neglected, the polarization
$\mathbf{P}(\mathbf{r})$ extends continuously up to the mathematical
dividing surface. If the dielectric borders vacuum, the polarization
field changes discontinuously from $\mathbf{P}(\mathbf{r}_S)$ at the
dielectric side of the surface to zero at its vacuum side.

The sum of the electric field from the surface charge density and the
external field is the Maxwell field $\mathbf{E}(\mathbf{r})$.  The
Maxwell field, as well as the polarization $\mathbf{P}(\mathbf{r})$,
cannot be directly measured, but can be retrieved from electric fields
inside cavities carved in the dielectric, as originally suggested by
Thompson and Maxwell.\cite{Thompson1872,Maxwell:V2} For the simplest
geometry of an empty spherical cavity, the solution of the Laplace
equation with Maxwell's boundary conditions gives the field at the
cavity center (cavity field)\cite{Landau8,Boettcher:73}
\begin{equation}
  \label{eq:1}
  E_{\text{M}} = \frac{3}{2\epsilon + 1} E_{\text{ext}}  , 
\end{equation}
where $\epsilon$ is the dielectric constant of the medium surrounding the
cavity.  The qualitative prediction of this result is that external
field $E_{\text{ext}}$ is diminished by a factor of $3/(2\epsilon)$
inside cavities created in highly polar dielectrics such as
water. This scenario then describes a well-defined physical setup
testable by laboratory or numerical experiment.

One wonders to what extent the mathematical formalism of Maxwell's
electrostatics applies to interfaces of polar liquids. The interface
is obviously not a mathematical surface, but has a finite width
(Laplace vs Poisson definition of the
interface\cite{NinhamNostro:10}). In addition, liquid dipoles have the
freedom to rotate and adjust to the lack of molecular interactions
from the cavity side of the interface. For polar liquids, this
restructuring results in preferential in-plane orientation of the
liquid dipoles at free planar and closed
interfaces.\cite{Lee:84,Lee:86,Valleau:87,Sokhan:97,Bratko:09,Rossky:10,Romero-Vargas-Castrillon:2011mz}
Unless the external field orients the dipoles off-plane, such
orientational structure eliminates the surface charge since $\sigma_P
= P_n \simeq 0$ (Fig.\ \ref{fig:1}b). The standard boundary conditions
of Maxwell's electrostatics do not apply, thus affecting the
observable cavity field.

Water presents a particularly important test case for understanding
the interfacial electrostatics of polar liquids. Strong hydrogen bonds
between surface waters can potentially prevent their reorientation not
only by a weak external field of the dielectric experiment, but also
by internal fields of solute charges.  Since it is the response of the
water dipoles that determines the free energy of
hydration,\cite{Pratt:94} this problem goes beyond the question of
measuring the electric field inside a dielectric cavity in Maxwell's
``gedanken experiment''.

We have previously approached the problem of liquid interfacial
electrostatics by literally following Maxwell's recipe and measuring,
by numerical simulations, the electric field inside an empty
hard-sphere cavity carved in the model fluid of dipolar hard
spheres.\cite{DMepl:08,DMjcp3:08} Theses studies have shown that the
cavity field indeed does not continuously decrease with increasing
$\epsilon$, as Eq.\ \eqref{eq:1} would suggest, but instead levels
off, as a function of $\epsilon$, at a value close to the result well
established in the theory of dielectrics. This limit corresponds to
the Lorentz virtual cavity.\cite{Boettcher:73} The latter is defined
as a part of the dielectric confined by a mathematical closed surface
inside it.  The Lorentz cavity field is then the field
produced by the dielectric outside this surface. Since no physical
interface is present, there is no physical polarization at the
interface and $\sigma_P=0$ by definition.\cite{com:Boettcher}

The electric field at the center of a virtual cavity is obtained by
integrating over the field contributions from homogeneous polarization
of the medium, instead of inhomogeneous polarization in the case of
Maxwell's electrostatics (see below).  The field accumulated by this
uniform polarization results in the Lorentz field at the cavity center
\begin{equation}
  \label{eq:2}
  E_{\text{L}} = \frac{\epsilon+2}{3\epsilon} E_{\text{ext}} .
\end{equation}
The main qualitative difference between this result and the standard
cavity field given by Eq.\ \eqref{eq:1} is that the  Lorentz field does
not decay to zero at $\epsilon\rightarrow\infty$ and instead levels
off at a non-zero value of $E_{\text{L}}/E_{\text{ext}}\rightarrow 1/3$.

The model dipolar system studied previously by us
\cite{DMepl:08,DMjcp3:08} has in principle shown that interfacial
orientational structure of a polar liquid might be closer to the
limiting scenario sketched in Fig.\ \ref{fig:1}b than to the standard,
solid-like picture sketched in Fig.\ \ref{fig:1}a.  The question of
practicality of this observation remains however open. A vacuum
hard-sphere cavity is a purely theoretical construct. Nevertheless,
the scenario arising from this model can potentially describe
non-polar particles solvated in real polar liquids. 

In order to approach this more realistic situation, we study here the
scenario that we have dubbed the ``Rossky cavity''. The actual system
is a Lennard-Jones (LJ) solute inserted in SPC/E water. As was
originally shown by Rossky and co-workers\cite{Lee:84} and supported
by many subsequent
studies,\cite{Lee:86,Valleau:87,Sokhan:97,Bratko:09,Rossky:10,Romero-Vargas-Castrillon:2011mz}
water dipoles orient in-plane at interfaces with non-polar
solutes. Although a non-polar solute studied here is certainly not a
cavity, the orientational water structure is highly resilient to
external perturbations and remains nearly intact for a broad range of
solute-solvent attractions.\cite{Torrie:93} The term ``cavity'' is
then used to stress that it is a large energy of the water hydrogen
bonds ($\Delta H \simeq 4-5$ $k_{\text{B}}T$ per bond for SPC/E
water\cite{Spoel:2006kx}) that supports the interfacial orientational
structure.  Therefore, in contrast to the virtual Lorentz cavity, this
physical cavity, like the traditional Maxwell cavity, is meant to
represent realistic measurements of local fields inside dielectrics,
in water in the present study. The Rossky cavity is meant to represent
physical situations when the interfacial structure is dominated by
in-plane orientations of water dipoles.  We show that the boundary
conditions imposed by molecular interfacial order of this cavity
produce the electrostatic response approaching the conditions expected
for the virtual Lorentz cavity and thus dramatically deviating from
the Maxwell cavity field.

\section{Results}
\label{sec:2}
\begin{figure}
  \centering
  \includegraphics*[width=8cm]{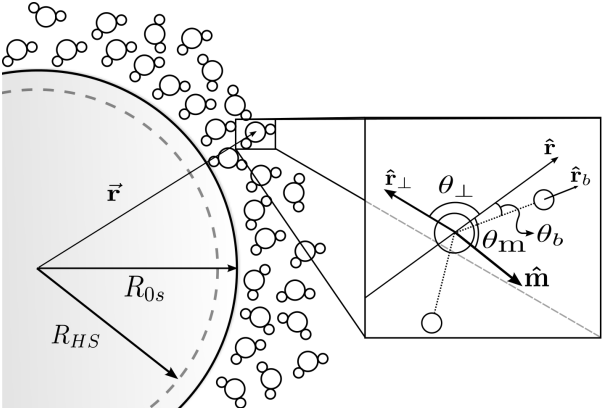}
  \caption{A cartoon of the Kihara solute in SPC/E water and the
    definition of the projection angles describing the
    orientations of waters in the first hydration layer in respect to
    the surface normal $\mathbf{\hat r} = -\mathbf{\hat n}$.  $\theta_b$ is the
    angle between O-H bond and $\mathbf{\hat r}$, $\theta_m$ is the
    angle between the water dipole and $\mathbf{\hat r}$,
    $\theta_{\bot}$ defines the orientation of the plane of
    $\mathrm{H_2O}$ (note that $\mathbf{\hat m} \cdot \mathbf{\hat
      r}_\bot = 0$).}
   \label{fig:2}
\end{figure}

The model considered here consists of a single non-polar solute
interacting with a large number of waters mimicking a typical
solvation experiment.  The solute-water interaction is modeled by a
Kihara potential combining a hard-sphere core of the radius
$R_{\text{HS}}$ with a surface LJ layer of width $\sigma_{0s}$ (Fig.\
\ref{fig:2})
\begin{equation}
  \label{eq:4}
  \phi_{0s}(r)  =  4 \epsilon_{0s} \left[ 
  \left( \frac{\sigma_{0s}}{r-R_{\text{HS}}} \right)^{12} - \left(
    \frac{\sigma_{0s}}{r-R_{\text{HS}}} \right)^{6} \right] .
\end{equation}
Here, ``0'' and ``s'' are used to label the solute and solvent
(water), respectively; $\epsilon_{0s}$ is the energy of solute-solvent
LJ attraction. Further, in order to map hard-sphere cavities used in
our previous studies\cite{DMepl:08,DMjcp3:08} on the ``soft'' Kihara
solute, we adopt the radius of the closest approach
\begin{equation}
\label{eq:5}
   R_{0s}  =  R_{\text{HS}} + \sigma_{0s} ,
\end{equation}
where $\sigma_{0s}=3.0$ \AA\ has been used in all numerical calculations.

The Kihara potential is the only solute-solvent interaction introduced
in our present model. We are therefore missing the effect of
electronic polarizability of the solute always present even for
non-polar particles. A crude way to estimate this effect is to notice
that only the ratio of the dielectric constants of two interfacing
media matters for the interfacial dielectric response in dielectric
theories.\cite{Landau8} While this prescription requires testing for
microscopic liquid interfaces, the model situation of a fully
non-polarizable solute can be mapped on real situations of solvated
non-polar particles by rescaling the solvent dielectric constant with
the dielectric constant $\epsilon_0$ of the solute,
$\epsilon\rightarrow \epsilon/ \epsilon_0$, if such property can be
reasonably defined. We will therefore proceed with the present model
using the term ``cavity field'' to describe the electric field at the
center of the solute.

Configurations of the water-solute mixture were produced by Molecular
Dynamics (MD) simulations of a single solute inserted in a box of SPC/E
waters at standard conditions (Supplementary Material
(SM)\cite{supplJCP}). The ratio of the cavity and external fields is
obtained in the linear response approximation as the correlator of the
electric field $\mathbf{E}_s$ produced by water at the solute center with
the water dipole moment $\mathbf{M}_s$
\begin{equation}
  \label{eq:3}
    E_c/E_{\text{ext}} =  1 + (\beta/3)\langle \delta {\bf E}_s \cdot \delta\mathbf{M}_s\rangle  -
  E_{\text{corr}}  .
\end{equation}
Here, $\delta {\bf E}_s$ and $\delta \mathbf{M}_s$ denote deviations
from the corresponding average values and $\beta = 1/(k_{\text{B}}T)$
is the inverse temperature. In addition, the correction term
$E_{\text{corr}}$ in the rhs of Eq.\ \eqref{eq:3} accounts for the
cutoff of the electrostatic interactions specific to a given
simulation protocol. The result for Ewald sums is given in the
SM,\cite{supplJCP} while the reaction field protocol was covered
in previous studies.\cite{DMepl:08,DMjcp3:08}

\begin{figure}
   \centering
   \includegraphics*[width=8cm]{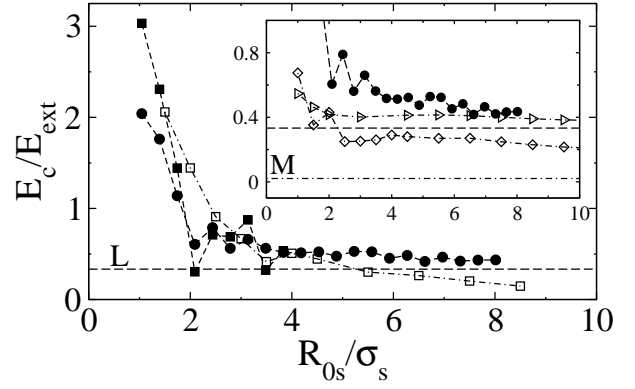}
   \caption{The cavity field inside Kihara solutes (filled points,
     connected by the dashed lines) compared to hard-sphere cavities
     inside dipolar fluids (open points, connected by the dash-dotted
     lines).\cite{DMepl:08} Results of MD simulations in the main
     panel refer to two values of the solute-solvent LJ attraction,
     $\epsilon_{0s}=0.65$ kJ/mol (circles) and 20.0 kJ/mol
     (squares). The results for the Kihara solutes are plotted against
     the radius of closest approach $R_{0s}/ \sigma_s$ defined by Eq.\
     \eqref{eq:5} ($\sigma_s=2.87$ \AA). Open points in the main panel
     and the inset refer to sets of data obtained for varied reduced
     dipole moment $(m^*)^2=\beta m^2/ \sigma_s^3$ of the dipolar
     hard-sphere fluids: 0.5 (diamonds), 1.0 (triangles), and 3.0
     (squares). The results for hard-sphere cavities in dipolar fluids
     are plotted against the radius of the closest hard-sphere
     solute-solvent approach $R_{0s}/ \sigma_s$; $\sigma_s$ is the
     diameter of the solvent hard spheres. The dashed horizontal lines
     in the main panel and in the inset refer to the Lorentz result
     (L) of Eq.\ \eqref{eq:2}. The dash-dotted horizontal line in the
     inset refers to the Maxwell result (M) of Eq.\ \eqref{eq:1}. }
   \label{fig:3}
\end{figure}

Equation \eqref{eq:3} was used to calculate the field inside the
solutes of varying size $R_{\text{HS}}$ of the Kihara hard-sphere
core, and the results are plotted in Fig.\ \ref{fig:3}. In order to
asses the effect of a uniform solute-solvent attraction on the cavity
field we have simulated the configurations at two values of the LJ
attraction, $\epsilon_{0s} = 0.65$ kJ/mol equal to the LJ energy
between oxygens of SPC/E water and $\epsilon_{0s} = 20$ kJ/mol close
to the energy required to break two hydrogen bonds in bulk SPC/E 
water.\cite{Spoel:2006kx} 

\begin{figure}
  \centering
  \includegraphics*[width=7cm]{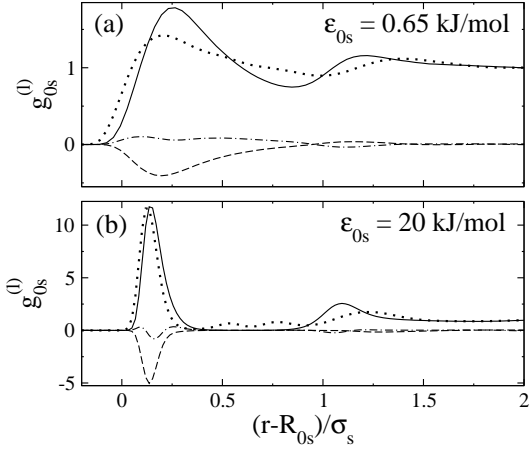}
  \caption{The solute-solvent distribution functions
    $g_{0s}^{(\ell)}(r)$ [Eq.\ \eqref{eq:8}] vs the distance from the
    surface of the Kihara solute ($R_{0s}=13$ \AA).  Shown are the solute-oxygen
    (solid lines) and solute-hydrogen (dotted lines) radial
    distribution functions ($\ell=0$) and orientational distributions
    for $\ell=1$ (dash-dotted lines) and $\ell=2$ (dashed lines); (a)
    refers to $\epsilon_{0s}=0.65~\textmd{kJ/mol}$ and (b) refers to
    $\epsilon_{0s}=20.0~\textmd{kJ/mol}$.}
   \label{fig:4}
\end{figure}

The smaller value of the LJ attraction perturbs little the water
structure at small values of $R_{\text{HS}}$, but results in a weakly
dewetted interface\cite{HummerPRL:98} at the end of the scale of
$R_{\text{HS}}$ values. This is meant to imply that the first peak of
the solute-solvent radial distribution function falls below the
corresponding peak of the solvent-solvent distribution function.  On
the contrary, the larger value of the solute-solvent LJ attraction
produced a substantial increase of the surface density of water as
judged from the solute-solvent pair distribution function (Fig.\
\ref{fig:4}). Furthermore, the radial distribution function is nearly
zero between the first and second hydration shells indicating layering
of water at the interface.\cite{BrovchenkoBook}

\begin{figure}
  \centering
  \includegraphics*[width=7cm]{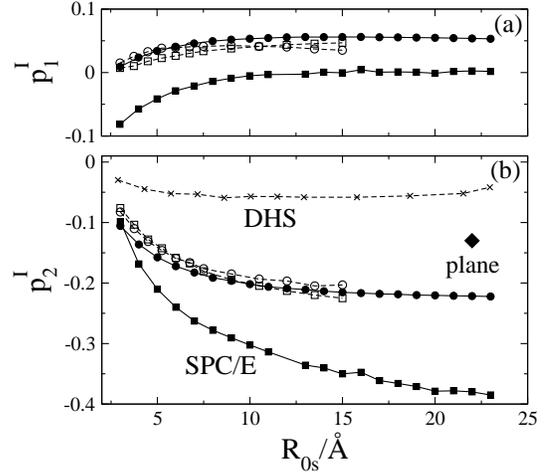}
  \caption{The first (upper panel) and second (lower panel)
    orientational order parameters $p_{1,2}^I$ of the first-shell
    SPC/E waters vs $R_{0s}=R_{\text{HS}} + \sigma_{0s} $.  The solid
    circles ($\epsilon_{0s}=0.65$ kJ/mol) and squares
    ($\epsilon_{0s}=20$ kJ/mol) refer to Kihara solutes in water at
    $T=300$ K. The corresponding open points refer to $T=273$ K. The
    crosses in the lower panel show $p_{2}^I$ for HS cavities in the
    fluid of dipolar hard spheres (DHS)\cite{DMpre1:08} with the
    reduced dipole moment $(m^*)^2=\beta m^2/ \sigma_s^3=3.0$; $m$ is
    the dipole moment and $\sigma_s$ is the HS diameter of the
    solvent.  The filled diamond labeled ``plane'' marks $p_2^I$ for a
    planar water interface from Ref.\ \onlinecite{Sokhan:97}. }
  \label{fig:5}
\end{figure}

In order to characterize the orientational structure of the
interfacial water dipoles, we define a series of distribution
functions recognizing the radial symmetry of the problem and producing
increasing symmetry orders of the water dipoles $\mathbf{\hat
  m}_j$ in projection on the outward normal to the spherical solute
surface $\mathbf{\hat r}_j$ at the position $\mathbf{r}_j$ of a water
molecule (oxygen coordinates are used for the center of mass). The
radial distribution functions are then defined in terms of
$\ell$-order Legendre polynomials $P_{\ell}(\cos \theta_{mj})$, $\cos
\theta_{mj} = \mathbf{\hat m}_j\cdot\mathbf{\hat r}_j$ as follows
\begin{equation}
  \label{eq:8}
  g_{0s}^{(\ell)} (r) = (V/N) \sum_j P_{\ell}(\cos \theta_{mj})
  \delta\left( \mathbf{r}_j - \mathbf{r} \right) .
\end{equation}
Here, $V$ and $N$ are the system volume and the number of particles,
respectively. The zeroth-order distribution function is then the
standard solute-oxygen radial distribution
$g_{0s}^{(0)}(r)=g_{0s}(r)$. In addition to these radial distributions
shown in Fig.\ \ref{fig:4}, we also calculate the order parameters of
the dipoles in the first hydration layer by integrating the radial
functions over the volume $V^I$ defined by the condition 
$0 < r <R_{0s}+\sigma_s/2$
\begin{equation}
  \label{eq:9}
  p^I_{\ell} = (N^I)^{-1} \rho_s \int_{V^I} g_{0s}^{(\ell)}(r) d\mathbf{r} .
\end{equation}
Here, $\rho_s$ is the water number density and the integrated radial
function is normalized to the number of waters in the first hydration
layer
\begin{equation}
  \label{eq:10}
  N^I = \rho_s \int_{V^I} g_{0s}^{(0)}(r)d\mathbf{r} .
\end{equation}
The results for first-order, $p^I_1$, and second-order, $p_2^I$, order
parameters vs the solute size are plotted in Fig.\ \ref{fig:5}. 

\begin{figure}
  \centering
  \includegraphics*[width=6cm]{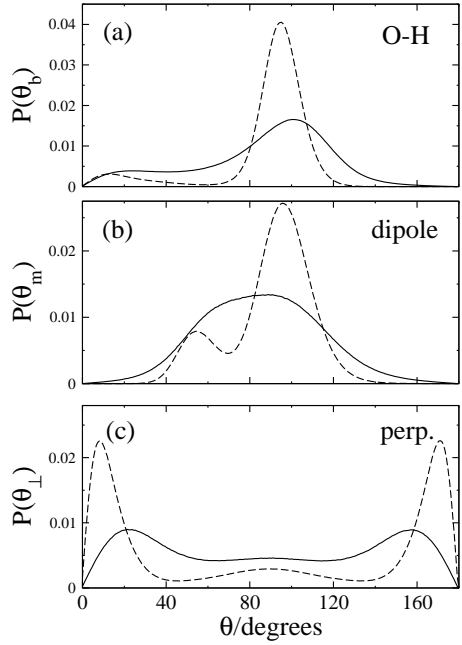}
  \caption{Distributions of three angles used to define the water
    orientations relative to the radial direction
    (Fig.\ \ref{fig:2}): (a) $\theta_b$ for the O--H bond ($\cos
    \theta_b = \hat{r}_b \cdot \hat{r}$), (b) $\theta_m$ for the water
    dipole ($\cos\theta_m =\hat{m} \cdot \hat{r}$), and (c)
    $\theta_{\bot}$ for the normal to $\mathrm{H_2O}$ plane
    ($\cos\theta_{\bot}=\hat{r}_\bot \cdot \hat{r}$); $R_{0s}=13$ \AA.  }
   \label{fig:6}
\end{figure}

The structure of surface waters is a ``squashed'' hexagonal ice
lattice in which two oxygen sublayers of hexagonal sheets are brought
into one plane of the closest solute-solvent approach corresponding to
the first peak of the solute-oxygen radial distribution function
(Fig.\ \ref{fig:4}). The hydrogens of first-shell waters are randomly
distributed for weak solute-solvent attraction of a hydrophobic
surface\cite{Lee:84} (Fig.\ \ref{fig:4}a), but get ordered with
increasing the attractive pull of the solute (Fig.\ \ref{fig:4}b). The
breaking of the sublayers of the hexagonal ice sheets results in
buckling of the O--H--O bond\cite{Sharp:10} from the straight line
found in hexagonal ice to preferential angles of $\sim 10^{\circ}$ (more
populated) and $\sim 70^{\circ}$ (less populated) (see the
SM\cite{supplJCP}). This distribution of hydrogen-bond angles is also
quite insensitive to the strength of the LJ attraction. Further, 
consistent with a broad distribution of the first-shell hydrogens
shown by the dotted lines in Fig.\ \ref{fig:4}, the water dipoles are
broadly distributed for weak solute-solvent attraction, but become
more ordered when LJ attraction becomes stronger (Fig.\
\ref{fig:6}b). In all cases, however, the in-plane orientation of the
water dipoles is preferred, and is highly resilient to the changes in
the solute-solvent attraction.
 
The distributions of projections of water O--H bonds and the normal to
the $\mathrm{H_2O}$ plane (Fig.\ \ref{fig:2}) are shown in Fig.\
\ref{fig:6}a,c. Both are consistent with the picture of the in-plane
orientation of the water dipoles, with the first-shell waters
populating variably the states of $\mathrm{H_2O}$ plane perpendicular
to the surface normal and slightly tilted, at $\sim 10^{\circ}$ in
respect to the normal. These two states are populated in the ratio of
about 1:2 for the weak attraction, but the tilted state becomes a
dominant one at the stronger attraction.  This observation is
consistent with the general tendency of interfacial waters to
increasingly occupy one particular orientational state with increasing
strength of attraction of either protons or oxygens to the solute.  In
contrast, the hydrogen disorder generally increases when a combination
of two types of attractions are present at the
interface,\cite{Shen:1998ys} a situation typical of hydrated proteins.

\section{Discussion}
The results for the electric field inside the Rossky cavity in water
are shown by filled points in Fig.\ \ref{fig:3}. They clearly tend to
the Lorentz field limit. We also plot in Fig.\ \ref{fig:3} (open
points) the results for the electric field inside hard-sphere cavities
in dipolar fluids.\cite{DMepl:08,DMjcp3:08} Although the two
situations are physically distinct, a non-polar solute in the former
case vs an actual cavity in the latter, the phenomenology of the polar
interfacial response is generic for both cases.  The common output of
these two sets of simulations suggests that the picture of
non-polarized interface represents the response of free and non-polar
interfaces of polar liquids in general.

It is instructive to see how the Maxwell cavity field and the Lorentz
field appear based on the assumptions regarding the surface charge
density (Fig.\ \ref{fig:1}a,b). We first assume a constant value for
the surface charge density $\sigma_P=\sigma_0$. The Maxwell electric
field in the medium is then the sum of the external field and the
radial field propagating from a uniformly charged interface
\begin{equation}
  \label{eq:12}
   \mathbf{E}(\mathbf{r}) = \frac{1}{\epsilon}\mathbf{E}_{\text{ext}}
   + \frac{4\pi R_0^2 \sigma_0}{\epsilon r^3} \mathbf{r} . 
\end{equation}
This electric field will polarize the dielectric yielding the
polarization field
$\mathbf{P}(\mathbf{r})=(\epsilon-1)\mathbf{E}(\mathbf{r})/(4\pi)$.
This polarization field is inhomogeneous close to the interface and
decays to the uniform polarization $\mathbf{P}$ in the bulk.  The
polarization of the dielectric will in turn produce its own electric
field, which, combined with the external field, gives the cavity field
\begin{equation}
  \label{eq:7}
  \mathbf{E}_c = \mathbf{E}_{\text{ext}} + \int_{\Omega}
  \mathbf{T}(\mathbf{r})\cdot \mathbf{P}(\mathbf{r}) d\mathbf{r} .
\end{equation}
Here, $\mathbf{T}(\mathbf{r})=\nabla \nabla r^{-1}$ is the dipolar
interaction tensor and the integral is over the dielectric occupying
the volume $\Omega$ outside the dielectric cavity of radius $R_0$.  The
radius $R_0$ does not need to be specified since the calculation
results do not depend on its value.

The radial polarization arising from the second summand in Eq.\
\eqref{eq:12} gives zero contribution to the cavity field and the
result of Eqs.\ \eqref{eq:12} and \eqref{eq:7} is the Lorentz field
given by Eq.\ \eqref{eq:2}. A constant, angular-independent surface
charge density makes therefore no contribution to the cavity field. It
however contributes to a non-zero, spatially constant electrostatic
potential inside the cavity.\cite{Ashbaugh:00}

Given axial symmetry of the problem, surface charge density can be
expanded in Legendre polynomials, $\sigma_P(\theta) =\sum_{\ell}
\sigma_{\ell} P_{\ell}(\cos \theta)$, where $\theta$ is the polar
angle with the direction of the external field. Since the zero-order term
$\sigma_0$ does not contribute to the cavity field, one can take the
first-order term and, neglecting the quadrupolar and higher moments, have
the dipolar approximation $\sigma_P(\theta)=\sigma_1\cos \theta$. The
dipole moment created by the solute interface is then
$M_0=\sigma_1\Omega_0$.  This is the situation sketched in Fig.\
\ref{fig:1}a.  The direct solution of the Laplace equation with
Maxwell's boundary conditions results in the interfacial dipole
\begin{equation}
  \label{eq:13}
  \mathbf{M}_0 = - \mathbf{P} \frac{3\Omega_0}{2\epsilon+1},
\end{equation}
where $\mathbf{P}=(\epsilon-1)/(4\pi\epsilon)\mathbf{E}_{\text{ext}}$.
The negative sign here indicates that the interface dipole
$\mathbf{M}_0$ orients oppositely to the external field, as is also
clear from Fig.\ \ref{fig:1}a.

\begin{figure}
  \centering
  \includegraphics*[width=8cm]{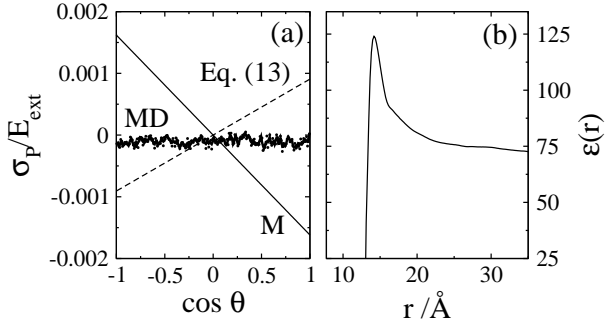}
  \caption{Surface charge density $\sigma_P(\theta)=P_n(\theta)$ vs
    the polar angle $\theta$ measured from the direction of the
    external field (a). The solid line (M) shows the result of
    Maxwell's electrostatics [Eq.\ \eqref{eq:11}, $\epsilon=72.2$ for
    SPC/E water at 300 K]. The dots refer to $\sigma_P$ calculated as
    a linear response to an external field from 40 ns of MD
    trajectories (see the SM\cite{supplJCP}). The MD $\sigma_P$ is
    collected from a water layer $R_{0s}\le r \le R_{0s} +
    0.1\sigma_s$ for the solute with the size $R_{0s}=13$ \AA\ and
    $\epsilon_{0s}=0.65$ kJ/mol. The dashed line shows the result of
    using $E_c$ from simulations to calculate $\chi_1$ from Eq.\
    \eqref{eq:14}. Panel (b) shows the profile of water's dielectric
    constant $\epsilon(r) = 1 +(4\pi\beta /3\Omega(r)) \langle \delta
    \mathbf{M}_s(r)\cdot \delta\mathbf{M}_s\rangle$ (see text). }
  \label{fig:7}
\end{figure}

The dipolar surface charge density following from Eq.\ \eqref{eq:13}
is shown by the solid line in Fig.\ \ref{fig:7}.  This surface charge
density creates a dipolar field, which adds to the external field to
produce the Maxwell field
\begin{equation}
  \label{eq:6}
  \mathbf{E}(\mathbf{r}) = \frac{1}{\epsilon}\mathbf{E}_{\text{ext}} +
   \mathbf{T}(\mathbf{r})\cdot \mathbf{M}_0 .
\end{equation}
When the polarization field $\mathbf{P}(\mathbf{r})$ is calculated
from Eq.\ \eqref{eq:6} and substituted into Eq.\ \eqref{eq:7}, 
the cavity field becomes
\begin{equation}
  \label{eq:14}
  \frac{E_c}{ E_{\text{ext}} } = \frac{\epsilon+2}{3\epsilon} +
  \frac{8\pi(\epsilon-1)}{9} \chi_1 ,
\end{equation}
where 
\begin{equation}
  \label{eq:18}
  \chi_1 =  \frac{M_0}{\Omega_0 E_{\text{ext}}}
\end{equation}
is the dipolar response function of the solute interface to the
polarizing external field.  The Maxwell result for $M_0$ in Eq.\
\eqref{eq:13} gives the standard cavity field of Eq.\ \eqref{eq:1},
while $\chi_1=M_0=0$ leads to the Lorentz field of Eq.\ \eqref{eq:2}.

Surface charge density $\sigma_P(\theta)$ provides a convenient
mathematical idealization of the polarization of the interface.  Its
direct calculation from real-space, finite-size simulations presents,
however, a significant challenge. The problem is illustrated in Fig.\
\ref{fig:7}a which shows $\sigma_P(\theta)$ calculated at the solute
surface from MD trajectories using the linear response approximation
(see the SM\cite{supplJCP}). The result is a clearly
angular-independent function, producing $\sigma_1\simeq 0$ and
$E_c\simeq E_{\text{L}}$. This outcome is, however, not entirely
consistent with a slightly positive $\sigma_1$ following from the
substitution of the simulated $E_c$ into Eq.\ \eqref{eq:14} (dashed
line in Fig.\ \ref{fig:7}a).

The reason of possible uncertainties in the attempts to calculate
$\sigma_P(\theta)$ from sampling surface dipole directions from MD
trajectories is illustrated in Fig.\ \ref{fig:7}b. The definition of a
mathematical dividing surface is uncertain when the dipolar response
is strongly varied at the interface, as is the case here. We show the
dielectric constant of the polar layer of radius $r$ from the cavity
center obtained from the linear response approximation
as\cite{DMcpl:11} $\epsilon(r) = 1 +(4\pi\beta/3\Omega(r)) \langle
\delta \mathbf{M}_s(r)\cdot \delta\mathbf{M}_s\rangle$, where
$\mathbf{M}_s(r)$ and $\Omega(r)$ are, respectively, the dipole moment
and volume of the solvent within the $r$-surface. The layer dielectric
constant peaks near the interface and then slowly decays to the bulk
dielectric constant $\epsilon$. Any surface drawn next to the
interface will therefore reflect a different variance of the surface
dipole, with a different outcome for $\sigma_P(\theta)$. How much the
functional form of $\sigma_P(\theta)$ changes depending on the surface
definition cannot be answered here. Attempts to obtain
$\sigma_P(\theta)$ at surfaces deeper into the bulk did not produce
converged functions and therefore are not shown here.

It seems worth emphasizing some critical differences between both
(Maxwell and Lorentz) continuum results and the molecular picture
offered by the present numerical simulations. The continuum cavity field
does not depend on the cavity size, but only on the cavity shape
(e.g., on the length/diameter ratio for a cylindrical
cavity\cite{Maxwell:V2,Boettcher:73}). This is of course the
consequence of neglecting the actual spread of the interfacial region
relative to the size of the solute, which is behind the definition of
the interface as a mathematical dividing surface. Our simulations
allow us to estimate the solute size at which this approximation
becomes valid. Figure \ref{fig:3} shows that the cavity field starts
leveling off to a size-independent limit at $R_{0s}/ \sigma_s \simeq
2-2.5$. It gives an estimate of the solute size at which the ``Lorentz
continuum'' starts to take hold: the solute needs to be $4-5$ times
larger than the solvent molecule. The ``Maxwell continuum'' is,
however, never reached in our simulations.

\section{Experimental observables}
The surface polarization effects discussed here have a number of
consequences observable in a macroscopic laboratory experiment. The total free
energy of polarizing the dielectric is measured by the dielectric
experiment and is affected by the dipole accumulated at the interface
with a non-polar solute. The total dipole of a mixture sample
$M_{\text{mix}}$ is reduced relative to the homogeneous solvent by the
volume excluded by the solute and, in addition, is affected by the dipole
of the interface
\begin{equation}
  \label{eq:15}
   \mathbf{M}_{\text{mix}} = \mathbf{P}\Omega - (2/3) (\epsilon-1) \mathbf{M}_0 N_0 . 
\end{equation}
Here, $N_0$ is the number of the solutes in the solution and $\Omega$,
as above, is the solvent volume. The solutes are assumed to be
non-interacting, although the theory can be extended to non-ideal
solutions requiring the solute-solute structure factor as an
additional input.\cite{DMpre:10}

The standard arguments of the theory of dielectrics then suggest the
equation for the dielectric constant of the
mixture\cite{DMjcp3:08,DMpre:10}
\begin{equation}
  \label{eq:11}
  \frac{\epsilon}{\epsilon_{\text{mix}}} = 1 + \eta_0 (\epsilon-1)(1 +
  (8\pi/3) \epsilon\chi_1), 
\end{equation}
where $\eta_0 = N_0\Omega_0/(N_0\Omega_0 + \Omega )$ is the volume
fraction of the solutes. Equations \eqref{eq:14} and \eqref{eq:11},
taken together, suggest that dielectric constants of low-concentration
solutions and cavity fields inside solutes both give experimental
routes to measure the dipolar response function of the interface
$\chi_1$. Combining two equations together, one gets
\begin{equation}
  \label{eq:21}
   \frac{\epsilon}{\epsilon_{\text{mix}}} = 1 + 3 \eta_0\left[\epsilon
     \frac{E_c}{E_{\text{ext}}} -1 \right]
\end{equation}

For the cavity field $E_c=E_{\text{M}}$ given by Maxwell's form in
Eq.\ \eqref{eq:1} one arrives at the result\cite{DMjcp3:08,DMpre:10}
\begin{equation}
  \label{eq:16}
  \frac{\epsilon}{\epsilon_{\text{mix}}} = 1 + \eta_0 \frac{3(\epsilon-1)}{2\epsilon+1}.
\end{equation}
This equation is consistent with the Maxwell-Wagner
theory\cite{Scaife:98} in the limit of small volume fraction $\eta_0$.

\begin{figure}
  \centering
  \includegraphics*[width=8cm]{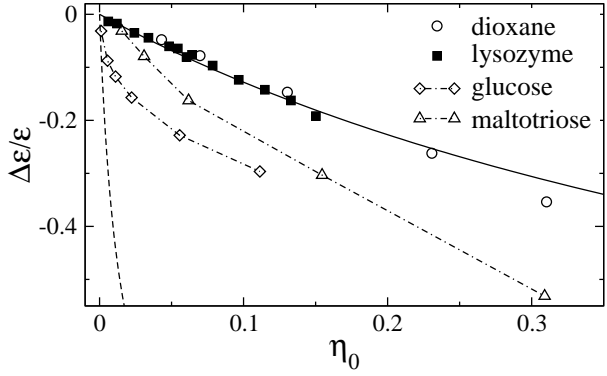}
  \caption{Relative dielectric constant increment $\Delta \epsilon /
    \epsilon$, $\Delta \epsilon= \epsilon_{\text{mix}}-\epsilon$ for
    several aqueous solutions: lysozyme\cite{Cametti:2011ys} (filled
    squares), dioxane\cite{Schrodle:2007kx} (open circles),
    glucose\cite{weingartner:1463} (open diamonds), and
    maltotriose\cite{weingartner:1463} (open triangles). The solid
    line (M) shows the result of Eq.\ \eqref{eq:16}, while the dashed
    line (L) is the dielectric increment for a solute with no surface
    charge density, $\sigma_1=0$, and therefore Lorentz result for the
    cavity field [Eq.\ \eqref{eq:14}]. Experimental and calculation
    results used to convert the experimentally reported solute
    concentrations to volume fractions can be found in the
    SM.\cite{supplJCP} The dash-dotted lines connect the experimental
    points for two saccharides. }
  \label{fig:8}
\end{figure}

Figure \ref{fig:8} illustrates the application of Eqs.\ \eqref{eq:11}
and \eqref{eq:16} to solutions of ionic and hydrogen-bonding
substances in
water.\cite{Cametti:2011ys,Schrodle:2007kx,weingartner:1463} The
closed squares in the figure show the relative dielectric constant
increment $\Delta \epsilon/ \epsilon$, $\Delta
\epsilon=\epsilon_{\text{mix}}-\epsilon$ for hydrated
lysozyme.\cite{Cametti:2011ys} At the pH of the measurements the
protein carries the total charge of $+10$ electron units. Open circles
in the figure show the results for the dioxane-water
mixture.\cite{Schrodle:2007kx} Dioxane offers strong hydrogen-bond
acceptor sites.  Finally, diamonds and triangles show, respectively,
hydrated glucose\cite{weingartner:1463} and
maltotriose,\cite{weingartner:1463} both offering multiple
hydrogen-bonding sites. The results for lysozyme and for two
saccharides refer to dielectric increments of the loss peak of the
water component of the solution.\cite{Cametti:2011ys,weingartner:1463}
This approach has allowed us to eliminate the contribution of the
permanent dipole moment of the solute, characterized by a much longer
relaxation time. The increment of the water loss peak 
reflects only the interfacial polarization described by the present
model. The dipole moment of dioxane is zero and the increment of the
static dielectric constant was used in that case.

The experimental results are compared to Maxwell [Eq.\ \eqref{eq:16},
solid line] and Lorentz [$\chi_1=0$ in Eq.\ \eqref{eq:11}, dashed
line] scenarios.  All solutes, potentially breaking the interfacial
network of hydrogen bonds, are either consistent with the Maxwell
interfacial polarization or fall between the Maxwell and Lorentz
predictions.

The Lorentz interface predicts a much enhanced sensitivity of the
dielectric constant to non-polar impurities, which also implies a
significant free energy penalty for polarizing such mixtures.
Experimental testing of the Lorentz scenario should therefore meet
with obvious solubility difficulties. However, scenarios intermediate
between the Lorentz and Maxwell limits clearly exist and the concept
of an effective surface charge density $\sigma_P$ provides a useful
conceptual framework for developing theories of polar response not
restricted to Maxwell's boundary conditions.

While uniform fields of the dielectric experiment measure the dipolar
response of the entire solution, non-uniform external fields give
access to the response function $\chi_1$ and therefore to the dipolar
polarization of the interface. If the external field varies on a scale
large compared to the dimension of the solute, the interfacial dipole
couples to the field gradient. The result is a force acting on the
solute
\begin{equation}
  \label{eq:17}
   F_z = (\Omega_0/2) \chi_1 \nabla_z E_{\text{ext}}(z)^2,  
\end{equation}
where the $z$-axis is chosen along the external field. This
phenomenon, known as dielectrophoresis,\cite{JonesBook:95} allows a
direct access to the polarization of the interface.  The Maxwell
solution for $\chi_1$ [Eqs.\ \eqref{eq:13} and \eqref{eq:18}] leads to
the standard dielectrophoresis coefficient $K \propto \chi_1$ used in
the theory of colloidal suspensions. \cite{JonesBook:95} In these
applications, $\epsilon$ is typically replaced with the ratio of the
dielectric constants of the solvent and the solute, $\epsilon
\rightarrow \epsilon/ \epsilon_0$, with the result
\begin{equation}
  \label{eq:20}
  K = \frac{\epsilon_0 - \epsilon}{\epsilon_0 + 2\epsilon}. 
\end{equation}
If the solute is more polar than the solvent, one gets positive
dielectrophoresis ($K>0$, $\chi_1 >0 $), and solute's attraction to a
stronger electric field. The opposite case of $K<0$ ($\chi_1<0$)
implies negative dielectrophoresis and thus repulsion of the solute
from the field.

\section{Summary}
\label{sec:3}
This paper extends our previous studies\cite{DMepl:08,DMjcp3:08} of
dipolar fluids interfacing hard-sphere cavities to attractive
non-polar solutes hydrated by SPC/E water. In-plane orientational
structure of the surface dipoles holds for interfaces of both dipolar
liquids and water. The orientational distribution of the interfacial
dipoles alters the boundary conditions of the polar response problem
monitored here in terms of the electric field inside the solute. Like
in previous studies of cavities in dipolar fluids, the field inside
the solute does not follow the prediction of Maxwell's electrostatics
[Eq.\ \eqref{eq:1}] and instead tends, with increasing solute size, to
the limit established by the Lorentz field of a non-polarized
interface [Eq.\ \eqref{eq:2}]. 

We find that the deviation of the cavity field from the Lorentz result
is given by the dipolar response function of the interface that also
enters the dielectric constant of an ideal solution. Finding the slope
of the dielectric increment of the low-concentration mixture vs the
solute concentration [Eqs.\ \eqref{eq:21} and \eqref{eq:11}] thus
provides a direct input into the cavity field [Eqs.\ \eqref{eq:21} and
\eqref{eq:14}]. This statement also applies to the frequency-dependent
response. As such, the dielectric increment of the fast water
component of the solution was taken in Fig.\ \ref{fig:8} to avoid the
effect of a slowly relaxing permanent dipole of the solute. Our
formalism is therefore applicable to ionic and polar solutes when the
frequency range exceeds the loss peak of the solute dipole. Along
these lines, the formalism can also be used at much higher frequencies
of UV/VIS light to find the refractive index corrections for rates of
radiative decay of photoexcited chromophores\cite{Toptygin:2003ly} and
quantum dots\cite{Wuister:2004zr} in solution. According to our
present results, this input should be sought from measuring the
refractive index of corresponding mixtures [Eq.\ \eqref{eq:21}].

We also find that small differences in the cavity field, which are
hard to resolve within the current simulation protocol, propagate in
very substantial differences in the dielectric response of a mixture. 
While this fact also spells out a thermodynamic difficulty of
preparing solutions that might test the Lorentz interface by
conventional dielectric spectroscopy, a large difference in the slopes
of dashed (Lorentz) and solid (Maxwell) lines in Fig.\ \ref{fig:8}
leaves much space to looking for intermediate scenarios. 

Returning to the question posed in the Introduction, the liquid
interface assumes the structure that makes its contribution to the
field inside and outside the solute essentially null. The whole
solvent response to an external polarizing field is given by the
uniform polarization of the bulk and does not include an inhomogeneous
component due to the surface charge [second summand in Eq.\
\eqref{eq:6}]. In retrospect, this outcome should have been
expected. The standard practice of liquid state theories and
corresponding numerical simulations suggests a short propagation
length of perturbations in liquids. A cavity or a non-polar solute
should therefore be ``invisible'' to an observable (e.g., Maxwell
field) measured a few solvent diameters from the interface, as indeed
our results show.  This result of course goes against the concept of
the surface charge inducing a long-ranged Coulomb perturbation in the
solvent.

The original concept of electric polarization envisioned by Maxwell
anticipates a limited, elastic response of medium's electric charge to
an external electric field.\cite{Maxwell:V2} For molecular dipoles of
a dielectric material, this concept implies limited small-amplitude
reorientations of the dipoles, aligning them with a weak external
field (linear response). This physical picture, which seems to match
the problem of dipolar polarization of a free solid-like interface,
was historically extended to liquid dielectrics which, in contrast,
have more extended ability to respond by both changing the positions
of their dipoles and producing large-amplitude dipolar
reorientations. In other words, the translational and rotational
mobility supported by the liquid phase allows the surface dipoles to
react to the creation of the interface by rearranging their dipolar
orientations in a way of diminishing the stress of a sharply varying
density profile.  This orientational order responds to a weak external
field by the rules that do not require a surface charge and the
induction of an interfacial dipole combining the negative and positive
lobes of the surface charge density (Fig.\ \ref{fig:1}a).  The
boundary conditions applicable to these systems differ from the ones
anticipated for Maxwell dielectrics.

By the way of yet another historical aberration, the standard theory
of dielectrics has been mostly applied not to free liquid interfaces,
for which Maxwell's construct was put forward, but to highly
hydrophilic and wetted interfaces covered with surface ions or to
solvation of molecular ions (Born theory of ion solvation and its
extensions). The agreement between observations done for these
surfaces and solutions with the predictions of dielectric models is
often used to support the conceptual framework of dielectric theories
developed for free or non-wetted (small attraction) surfaces. Our
present development cautions against this inconsistency and points out
that the orientational structure of the interface defines boundary
conditions of dielectric theories and with that the microscopic
and macroscopic fields observed near dielectric interfaces.

The picture of in-plane dipolar polarization of the liquid interface
is rather robust and insensitive to the interface curvature.  It holds
even for planar surfaces (filled diamond in Fig.\
\ref{fig:5})\cite{Sokhan:97,Shen:2006fr} suggesting that the
conclusions reached here for an admittedly narrow range of solute
radii may extend to larger solutes of potentially meso-to-macroscopic
size.

What matters for the boundary conditions entering the electrostatic
response functions is the orientational distribution of surface
waters. This can be altered by surface ions and polar
groups\cite{Verdaguer:2006ul,Barnette:2008pd,Jena:2010qy,Romero-Vargas-Castrillon:2011mz}
which potentially can create a non-zero surface charge density
$\sigma_P\ne 0$ matching the standard conditions of Maxwell's
electrostatics. The standard prescriptions will apply to those
cases. It follows directly from Coulomb's law that the change of the
normal component of the electric field at the interface of a
dielectric with vacuum is related to the surface charge density,
$-\Delta E_n = 4\pi \sigma_P$.  The near-zero $\sigma_P$ then implies
the continuation of the normal component of the electric field, and,
therefore, the continuation of the vector $\mathbf{E}$ of the Maxwell
field across a dielectric interface.

Liquids with large cohesive energy, network liquids, such as water, in
particular, seem to be particularly relevant to this discussion.  The
strength of hydrogen bonds ($\sim 4-5$ $k_{\text{B}}T$ per
bond\cite{Spoel:2006kx}) is so significant that in-plane dipolar
pattern may withstand local electric fields of solute partial
charges. The ratio of the characteristic strengths of the
solute-solvent to solvent-solvent interactions will therefore
determine the orientational distribution of the surface dipoles and,
ultimately, the type of boundary conditions used in calculations of
the electrostatic response.

\acknowledgments This research was supported by the National Science
Foundation (CHE-0910905).  CPU time was provided by the National
Science Foundation through TeraGrid resources (TG-MCB080116N).  We are
grateful to Peter Rossky for useful discussions and comments on the
manuscript.

%


\begin{thebibliography}{10}%
\makeatletter
\providecommand \@ifxundefined [1]{%
 \ifx #1\undefined \expandafter \@firstoftwo
 \else \expandafter \@secondoftwo
\fi
}%
\providecommand \@ifnum [1]{%
 \ifnum #1\expandafter \@firstoftwo
 \else \expandafter \@secondoftwo
\fi
}%
\providecommand \enquote [1]{``#1''}%
\providecommand \bibnamefont  [1]{#1}%
\providecommand \bibfnamefont [1]{#1}%
\providecommand \citenamefont [1]{#1}%
\providecommand\href[0]{\@sanitize\@href}%
\providecommand\@href[1]{\endgroup\@@startlink{#1}\endgroup\@@href}%
\providecommand\@@href[1]{#1\@@endlink}%
\providecommand \@sanitize [0]{\begingroup\catcode`\&12\catcode`\#12\relax}%
\@ifxundefined \pdfoutput {\@firstoftwo}{%
 \@ifnum{\z@=\pdfoutput}{\@firstoftwo}{\@secondoftwo}%
}{%
 \providecommand\@@startlink[1]{\leavevmode}%
 \providecommand\@@endlink[0]{}%
}{%
 \providecommand\@@startlink[1]{%
  \leavevmode
  \pdfstartlink
   attr{/Border[0 0 1 ]/H/I/C[0 1 1]}%
   user{/Subtype/Link/A<</Type/Action/S/URI/URI(#1)>>}%
  \relax
 }%
 \providecommand\@@endlink[0]{\pdfendlink}%
}%
\providecommand \url  [0]{\begingroup\@sanitize \@url }%
\providecommand \@url [1]{\endgroup\@href {#1}{\urlprefix}}%
\providecommand \urlprefix [0]{URL }%
\providecommand \Eprint[0]{\href }%
\@ifxundefined \urlstyle {%
  \providecommand \doi [1]{doi:\discretionary{}{}{}#1}%
}{%
  \providecommand \doi [0]{doi:\discretionary{}{}{}\begingroup
  \urlstyle{rm}\Url }%
}%
\providecommand \doibase [0]{http://dx.doi.org/}%
\providecommand \Doi[1]{\href{\doibase#1}}%
\providecommand \selectlanguage [0]{\@gobble}%
\providecommand \bibinfo [0]{\@secondoftwo}%
\providecommand \bibfield [0]{\@secondoftwo}%
\providecommand \translation [1]{[#1]}%
\providecommand \BibitemOpen[0]{}%
\providecommand \bibitemStop [0]{}%
\providecommand \bibitemNoStop [0]{.\EOS\space}%
\providecommand \EOS [0]{\spacefactor3000\relax}%
\providecommand \BibitemShut [1]{\csname bibitem#1\endcsname}%
\bibitem{Landau8}%
  \BibitemOpen
  \bibfield{author}{%
  \bibinfo {author} {\bibfnamefont{L.~D.}\ \bibnamefont{Landau}}\ and\ \bibinfo
  {author} {\bibfnamefont{E.~M.}\ \bibnamefont{Lifshitz}},\ }%
  \emph{\bibinfo {title} {Electrodynamics of continuous media}}\ (\bibinfo
  {publisher} {Pergamon},\ \bibinfo {address} {Oxford},\ \bibinfo {year}
  {1984})\BibitemShut{NoStop}%
\bibitem{Thompson1872}%
  \BibitemOpen
  \bibfield{author}{%
  \bibinfo {author} {\bibfnamefont{W.}~\bibnamefont{{Thompson Lord Kelvin}}},\
  }%
  \emph{\bibinfo {title} {Reprint of Papers on Electrostatics and Magnetism}},\
  \bibinfo {edition} {2nd}\ ed.\ (\bibinfo {publisher} {MacMillan and Co.},\
  \bibinfo {address} {London},\ \bibinfo {year} {1884, sec
  479})\BibitemShut{NoStop}%
\bibitem{Maxwell:V2}%
  \BibitemOpen
  \bibfield{author}{%
  \bibinfo {author} {\bibfnamefont{J.~C.}\ \bibnamefont{Maxwell}},\ }%
  \emph{\bibinfo {title} {A Treatise on Electricity and Magnetism}},\
  Vol.~\bibinfo {volume} {2}\ (\bibinfo {publisher} {Dover Publications},\
  \bibinfo {address} {New York},\ \bibinfo {year} {1954, secs.
  395-400})\BibitemShut{NoStop}%
\bibitem{Boettcher:73}%
  \BibitemOpen
  \bibfield{author}{%
  \bibinfo {author} {\bibfnamefont{C.~J.~F.}\ \bibnamefont{B{{\"o}}ttcher}},\
  }%
  \emph{\bibinfo {title} {Theory of Electric Polarization}},\ Vol.~\bibinfo
  {volume} {1}\ (\bibinfo {publisher} {Elsevier},\ \bibinfo {address}
  {Amsterdam},\ \bibinfo {year} {1973})\BibitemShut{NoStop}%
\bibitem{NinhamNostro:10}%
  \BibitemOpen
  \bibfield{author}{%
  \bibinfo {author} {\bibfnamefont{B.~W.}\ \bibnamefont{Ninham}}\ and\ \bibinfo
  {author} {\bibfnamefont{P.}~\bibnamefont{{Lo Nostro}}},\ }%
  \emph{\bibinfo {title} {Molecular Forces and Self Assembly In Colloids, Nano
  Sciences, and Biology}}\ (\bibinfo {publisher} {Cambridge University Press},\
  \bibinfo {address} {Cambridge},\ \bibinfo {year} {2010})\BibitemShut{NoStop}%
\bibitem{Lee:84}%
  \BibitemOpen
  \bibfield{author}{%
  \bibinfo {author} {\bibfnamefont{C.~Y.}\ \bibnamefont{Lee}}, \bibinfo
  {author} {\bibfnamefont{J.~A.}\ \bibnamefont{McCammon}},\ and\ \bibinfo
  {author} {\bibfnamefont{P.~J.}\ \bibnamefont{Rossky}},\ }%
  \bibfield{journal}{%
  \bibinfo {journal} {J.\ Chem.\ Phys.}\ }%
  \textbf{\bibinfo {volume} {80}},\ \bibinfo {pages} {4448} (\bibinfo {year}
  {1984})\BibitemShut{NoStop}%
\bibitem{Lee:86}%
  \BibitemOpen
  \bibfield{author}{%
  \bibinfo {author} {\bibfnamefont{S.~H.}\ \bibnamefont{Lee}}, \bibinfo
  {author} {\bibfnamefont{J.~C.}\ \bibnamefont{Rasaiah}},\ and\ \bibinfo
  {author} {\bibfnamefont{J.~B.}\ \bibnamefont{Hubbard}},\ }%
  \bibfield{journal}{%
  \bibinfo {journal} {J. Chem. Phys.}\ }%
  \textbf{\bibinfo {volume} {85}},\ \bibinfo {pages} {5232} (\bibinfo {year}
  {1986})\BibitemShut{NoStop}%
\bibitem{Valleau:87}%
  \BibitemOpen
  \bibfield{author}{%
  \bibinfo {author} {\bibfnamefont{J.~P.}\ \bibnamefont{Valleau}}\ and\
  \bibinfo {author} {\bibfnamefont{A.~A.}\ \bibnamefont{Gardner}},\ }%
  \bibfield{journal}{%
  \bibinfo {journal} {J. Chem. Phys.}\ }%
  \textbf{\bibinfo {volume} {86}},\ \bibinfo {pages} {4162} (\bibinfo {year}
  {1987})\BibitemShut{NoStop}%
\bibitem{Sokhan:97}%
  \BibitemOpen
  \bibfield{author}{%
  \bibinfo {author} {\bibfnamefont{V.~P.}\ \bibnamefont{Sokhan}}\ and\ \bibinfo
  {author} {\bibfnamefont{D.~J.}\ \bibnamefont{Tildesley}},\ }%
  \bibfield{journal}{%
  \bibinfo {journal} {Mol. Phys.}\ }%
  \textbf{\bibinfo {volume} {92}},\ \bibinfo {pages} {625} (\bibinfo {year}
  {1997})\BibitemShut{NoStop}%
\bibitem{Bratko:09}%
  \BibitemOpen
  \bibfield{author}{%
  \bibinfo {author} {\bibfnamefont{D.}~\bibnamefont{Bratko}}, \bibinfo {author}
  {\bibfnamefont{C.~D.}\ \bibnamefont{Daub}},\ and\ \bibinfo {author}
  {\bibfnamefont{A.}~\bibnamefont{Luzar}},\ }%
  \bibfield{journal}{%
  \bibinfo {journal} {Faraday Disc.}\ }%
  \textbf{\bibinfo {volume} {141}},\ \bibinfo {pages} {55} (\bibinfo {year}
  {2009})\BibitemShut{NoStop}%
\bibitem{Rossky:10}%
  \BibitemOpen
  \bibfield{author}{%
  \bibinfo {author} {\bibfnamefont{P.~J.}\ \bibnamefont{Rossky}},\ }%
  \bibfield{journal}{%
  \bibinfo {journal} {Farad. Disc.}\ }%
  \textbf{\bibinfo {volume} {146}},\ \bibinfo {pages} {13} (\bibinfo {year}
  {2010})\BibitemShut{NoStop}%
\bibitem{Romero-Vargas-Castrillon:2011mz}%
  \BibitemOpen
  \bibfield{author}{%
  \bibinfo {author}
  {\bibfnamefont{S.}~\bibnamefont{Romero-Vargas~Castrill{\'o}n}}, \bibinfo
  {author} {\bibfnamefont{N.}~\bibnamefont{Giovambattista}}, \bibinfo {author}
  {\bibfnamefont{I.~A.}\ \bibnamefont{Aksay}},\ and\ \bibinfo {author}
  {\bibfnamefont{P.~G.}\ \bibnamefont{Debenedetti}},\ }%
  \bibfield{journal}{%
  \bibinfo {journal} {J. Phys. Chem. C}\ }%
  \textbf{\bibinfo {volume} {115}},\ \bibinfo {pages} {4624} (\bibinfo {year}
  {2011})\BibitemShut{NoStop}%
\bibitem{Pratt:94}%
  \BibitemOpen
  \bibfield{author}{%
  \bibinfo {author} {\bibfnamefont{L.~R.}\ \bibnamefont{Pratt}}, \bibinfo
  {author} {\bibfnamefont{G.}~\bibnamefont{Hummer}},\ and\ \bibinfo {author}
  {\bibfnamefont{A.~E.}\ \bibnamefont{Garcia}},\ }%
  \bibfield{journal}{%
  \bibinfo {journal} {Biophys. Chem.}\ }%
  \textbf{\bibinfo {volume} {51}},\ \bibinfo {pages} {147} (\bibinfo {year}
  {1994})\BibitemShut{NoStop}%
\bibitem{DMepl:08}%
  \BibitemOpen
  \bibfield{author}{%
  \bibinfo {author} {\bibfnamefont{D.~R.}\ \bibnamefont{Martin}}\ and\ \bibinfo
  {author} {\bibfnamefont{D.~V.}\ \bibnamefont{Matyushov}},\ }%
  \bibfield{journal}{%
  \bibinfo {journal} {Europhys.\ Lett.}\ }%
  \textbf{\bibinfo {volume} {82}},\ \bibinfo {pages} {16003} (\bibinfo {year}
  {2008})\BibitemShut{NoStop}%
\bibitem{DMjcp3:08}%
  \BibitemOpen
  \bibfield{author}{%
  \bibinfo {author} {\bibfnamefont{D.~R.}\ \bibnamefont{Martin}}\ and\ \bibinfo
  {author} {\bibfnamefont{D.~V.}\ \bibnamefont{Matyushov}},\ }%
  \bibfield{journal}{%
  \bibinfo {journal} {J.\ Chem.\ Phys.}\ }%
  \textbf{\bibinfo {volume} {129}},\ \bibinfo {pages} {174508} (\bibinfo {year}
  {2008})\BibitemShut{NoStop}%
\bibitem{com:Boettcher}%
  \BibitemOpen
  \bibinfo {note} {The concept of surface charge density in used in Ref.\
  \onlinecite{Boettcher:73} to calculate the Lorentz field. While the procedure
  yields the correct algebraic result, the usage of this concept for a virtual
  surface is incorrect and should be avoided.}\BibitemShut{Stop}%
\bibitem{Torrie:93}%
  \BibitemOpen
  \bibfield{author}{%
  \bibinfo {author} {\bibnamefont{G.M.Torrie}}\ and\ \bibinfo {author}
  {\bibfnamefont{G.~N.}\ \bibnamefont{Patey}},\ }%
  \bibfield{journal}{%
  \bibinfo {journal} {J. Phys. Chem.}\ }%
  \textbf{\bibinfo {volume} {97}},\ \bibinfo {pages} {12909} (\bibinfo {year}
  {1993})\BibitemShut{NoStop}%
\bibitem{Spoel:2006kx}%
  \BibitemOpen
  \bibfield{author}{%
  \bibinfo {author} {\bibfnamefont{D.}~\bibnamefont{van~der Spoel}}, \bibinfo
  {author} {\bibfnamefont{P.~J.}\ \bibnamefont{van Maaren}}, \bibinfo {author}
  {\bibfnamefont{P.}~\bibnamefont{Larsson}},\ and\ \bibinfo {author}
  {\bibfnamefont{N.}~\bibnamefont{T{\^\i}mneanu}},\ }%
  \bibfield{journal}{%
  \bibinfo {journal} {J. Phys. Chem. B}\ }%
  \textbf{\bibinfo {volume} {110}},\ \bibinfo {pages} {4393} (\bibinfo {year}
  {2006})\BibitemShut{NoStop}%
\bibitem{supplJCP}%
  \BibitemOpen
  \bibinfo {note} {See supplementary material at [URL will be inserted by AIP]
  for details of the simulation protocol}\BibitemShut{NoStop}%
\bibitem{HummerPRL:98}%
  \BibitemOpen
  \bibfield{author}{%
  \bibinfo {author} {\bibfnamefont{G.}~\bibnamefont{Hummer}}\ and\ \bibinfo
  {author} {\bibfnamefont{S.}~\bibnamefont{Garde}},\ }%
  \bibfield{journal}{%
  \bibinfo {journal} {Phys. Rev. Lett.}\ }%
  \textbf{\bibinfo {volume} {80}},\ \bibinfo {pages} {4193} (\bibinfo {year}
  {1998})\BibitemShut{NoStop}%
\bibitem{BrovchenkoBook}%
  \BibitemOpen
  \bibfield{author}{%
  \bibinfo {author} {\bibfnamefont{I.}~\bibnamefont{Brovchenko}}\ and\ \bibinfo
  {author} {\bibfnamefont{A.}~\bibnamefont{Oleinikova}},\ }%
  \emph{\bibinfo {title} {Interfacial and confined water}}\ (\bibinfo
  {publisher} {Elsevier},\ \bibinfo {address} {Amsterdam},\ \bibinfo {year}
  {2008})\BibitemShut{NoStop}%
\bibitem{DMpre1:08}%
  \BibitemOpen
  \bibfield{author}{%
  \bibinfo {author} {\bibfnamefont{D.~R.}\ \bibnamefont{Martin}}\ and\ \bibinfo
  {author} {\bibfnamefont{D.~V.}\ \bibnamefont{Matyushov}},\ }%
  \bibfield{journal}{%
  \bibinfo {journal} {Phys. Rev. E}\ }%
  \textbf{\bibinfo {volume} {78}},\ \bibinfo {pages} {041206} (\bibinfo {year}
  {2008})\BibitemShut{NoStop}%
\bibitem{Sharp:10}%
  \BibitemOpen
  \bibfield{author}{%
  \bibinfo {author} {\bibfnamefont{K.~A.}\ \bibnamefont{Sharp}}\ and\ \bibinfo
  {author} {\bibfnamefont{J.~M.}\ \bibnamefont{Vanderkooi}},\ }%
  \bibfield{journal}{%
  \bibinfo {journal} {Acc.\ Chem.\ Res.}\ }%
  \textbf{\bibinfo {volume} {43}},\ \bibinfo {pages} {231} (\bibinfo {year}
  {2010})\BibitemShut{NoStop}%
\bibitem{Shen:1998ys}%
  \BibitemOpen
  \bibfield{author}{%
  \bibinfo {author} {\bibfnamefont{Y.~R.}\ \bibnamefont{Shen}},\ }%
  \bibfield{journal}{%
  \bibinfo {journal} {Solid State Comm.}\ }%
  \textbf{\bibinfo {volume} {108}},\ \bibinfo {pages} {399} (\bibinfo {year}
  {1998})\BibitemShut{NoStop}%
\bibitem{Ashbaugh:00}%
  \BibitemOpen
  \bibfield{author}{%
  \bibinfo {author} {\bibfnamefont{H.~S.}\ \bibnamefont{Ashbaugh}},\ }%
  \bibfield{journal}{%
  \bibinfo {journal} {J.\ Phys.\ Chem. B}\ }%
  \textbf{\bibinfo {volume} {104}},\ \bibinfo {pages} {7235} (\bibinfo {year}
  {2000})\BibitemShut{NoStop}%
\bibitem{DMcpl:11}%
  \BibitemOpen
  \bibfield{author}{%
  \bibinfo {author} {\bibfnamefont{A.~D.}\ \bibnamefont{Friesen}}\ and\
  \bibinfo {author} {\bibfnamefont{D.~V.}\ \bibnamefont{Matyushov}},\ }%
  \bibfield{journal}{%
  \bibinfo {journal} {Chem. Phys. Lett.},\ \bibinfo {pages}
  {10.1016/j.cplett.2011.06.031}}%
   (\bibinfo {year} {2011})\BibitemShut{NoStop}%
\bibitem{DMpre:10}%
  \BibitemOpen
  \bibfield{author}{%
  \bibinfo {author} {\bibfnamefont{D.~V.}\ \bibnamefont{Matyushov}},\ }%
  \bibfield{journal}{%
  \bibinfo {journal} {Phys. Rev. E}\ }%
  \textbf{\bibinfo {volume} {81}},\ \bibinfo {pages} {021914} (\bibinfo {year}
  {2010})\BibitemShut{NoStop}%
\bibitem{Scaife:98}%
  \BibitemOpen
  \bibfield{author}{%
  \bibinfo {author} {\bibfnamefont{B.~K.~P.}\ \bibnamefont{Scaife}},\ }%
  \emph{\bibinfo {title} {Principles of dielectrics}}\ (\bibinfo {publisher}
  {Clarendon Press},\ \bibinfo {address} {Oxford},\ \bibinfo {year}
  {1998})\BibitemShut{NoStop}%
\bibitem{Cametti:2011ys}%
  \BibitemOpen
  \bibfield{author}{%
  \bibinfo {author} {\bibfnamefont{C.}~\bibnamefont{Cametti}}, \bibinfo
  {author} {\bibfnamefont{S.}~\bibnamefont{Marchetti}}, \bibinfo {author}
  {\bibfnamefont{C.~M.~C.}\ \bibnamefont{Gambi}},\ and\ \bibinfo {author}
  {\bibfnamefont{G.}~\bibnamefont{Onori}},\ }%
  \bibfield{journal}{%
  \bibinfo {journal} {J. Phys. Chem. B}\ }%
  \textbf{\bibinfo {volume} {115}},\ \bibinfo {pages} {7144} (\bibinfo {year}
  {2011})\BibitemShut{NoStop}%
\bibitem{Schrodle:2007kx}%
  \BibitemOpen
  \bibfield{author}{%
  \bibinfo {author} {\bibfnamefont{S.}~\bibnamefont{Schr{\"o}dle}}, \bibinfo
  {author} {\bibfnamefont{G.}~\bibnamefont{Hefter}},\ and\ \bibinfo {author}
  {\bibfnamefont{R.}~\bibnamefont{Buchner}},\ }%
  \bibfield{journal}{%
  \bibinfo {journal} {J. Phys. Chem. B}\ }%
  \textbf{\bibinfo {volume} {111}},\ \bibinfo {pages} {5946} (\bibinfo {year}
  {2007})\BibitemShut{NoStop}%
\bibitem{weingartner:1463}%
  \BibitemOpen
  \bibfield{author}{%
  \bibinfo {author} {\bibfnamefont{H.}~\bibnamefont{Weing{\"a}rtner}}, \bibinfo
  {author} {\bibfnamefont{A.}~\bibnamefont{Knocks}}, \bibinfo {author}
  {\bibfnamefont{S.}~\bibnamefont{Boresch}}, \bibinfo {author}
  {\bibfnamefont{P.}~\bibnamefont{Hochtl}},\ and\ \bibinfo {author}
  {\bibfnamefont{O.}~\bibnamefont{Steinhauser}},\ }%
  \bibfield{journal}{%
  \bibinfo {journal} {J. Chem. Phys.}\ }%
  \textbf{\bibinfo {volume} {115}},\ \bibinfo {pages} {1463} (\bibinfo {year}
  {2001})\BibitemShut{NoStop}%
\bibitem{JonesBook:95}%
  \BibitemOpen
  \bibfield{author}{%
  \bibinfo {author} {\bibfnamefont{T.~B.}\ \bibnamefont{Jones}},\ }%
  \emph{\bibinfo {title} {Electromechanics of Particles}}\ (\bibinfo
  {publisher} {Cambridge University Press},\ \bibinfo {address} {Cambridge},\
  \bibinfo {year} {1995})\BibitemShut{NoStop}%
\bibitem{Toptygin:2003ly}%
  \BibitemOpen
  \bibfield{author}{%
  \bibinfo {author} {\bibfnamefont{D.}~\bibnamefont{Toptygin}},\ }%
  \bibfield{journal}{%
  \bibinfo {journal} {J. Fluoresc.}\ }%
  \textbf{\bibinfo {volume} {13}},\ \bibinfo {pages} {201} (\bibinfo {year}
  {2003})\BibitemShut{NoStop}%
\bibitem{Wuister:2004zr}%
  \BibitemOpen
  \bibfield{author}{%
  \bibinfo {author} {\bibfnamefont{S.~F.}\ \bibnamefont{Wuister}}, \bibinfo
  {author} {\bibfnamefont{C.~D.}\ \bibnamefont{Donega}},\ and\ \bibinfo
  {author} {\bibfnamefont{A.}~\bibnamefont{Meijerink}},\ }%
  \bibfield{journal}{%
  \bibinfo {journal} {J. Chem. Phys.}\ }%
  \textbf{\bibinfo {volume} {121}},\ \bibinfo {pages} {4310} (\bibinfo {year}
  {2004})\BibitemShut{NoStop}%
\bibitem{Shen:2006fr}%
  \BibitemOpen
  \bibfield{author}{%
  \bibinfo {author} {\bibfnamefont{Y.~R.}\ \bibnamefont{Shen}}\ and\ \bibinfo
  {author} {\bibfnamefont{V.}~\bibnamefont{Ostroverkhov}},\ }%
  \bibfield{journal}{%
  \bibinfo {journal} {Chem. Rev.}\ }%
  \textbf{\bibinfo {volume} {106}},\ \bibinfo {pages} {1140} (\bibinfo {year}
  {2006})\BibitemShut{NoStop}%
\bibitem{Verdaguer:2006ul}%
  \BibitemOpen
  \bibfield{author}{%
  \bibinfo {author} {\bibfnamefont{A.}~\bibnamefont{Verdaguer}}, \bibinfo
  {author} {\bibfnamefont{G.~M.}\ \bibnamefont{Sacha}}, \bibinfo {author}
  {\bibfnamefont{H.}~\bibnamefont{Bluhm}},\ and\ \bibinfo {author}
  {\bibfnamefont{M.}~\bibnamefont{Salmeron}},\ }%
  \bibfield{journal}{%
  \bibinfo {journal} {Chem. Rev.}\ }%
  \textbf{\bibinfo {volume} {106}},\ \bibinfo {pages} {1478} (\bibinfo {year}
  {2006})\BibitemShut{NoStop}%
\bibitem{Barnette:2008pd}%
  \BibitemOpen
  \bibfield{author}{%
  \bibinfo {author} {\bibfnamefont{A.~L.}\ \bibnamefont{Barnette}}, \bibinfo
  {author} {\bibfnamefont{D.~B.}\ \bibnamefont{Asay}},\ and\ \bibinfo {author}
  {\bibfnamefont{S.~H.}\ \bibnamefont{Kim}},\ }%
  \bibfield{journal}{%
  \bibinfo {journal} {Phys. Chem. Chem. Phys.}\ }%
  \textbf{\bibinfo {volume} {10}},\ \bibinfo {pages} {4981} (\bibinfo {year}
  {2008})\BibitemShut{NoStop}%
\bibitem{Jena:2010qy}%
  \BibitemOpen
  \bibfield{author}{%
  \bibinfo {author} {\bibfnamefont{K.~C.}\ \bibnamefont{Jena}}\ and\ \bibinfo
  {author} {\bibfnamefont{D.~K.}\ \bibnamefont{Hore}},\ }%
  \bibfield{journal}{%
  \bibinfo {journal} {Phys. Chem. Chem. Phys.}\ }%
  \textbf{\bibinfo {volume} {12}},\ \bibinfo {pages} {14383} (\bibinfo {year}
  {2010})\BibitemShut{NoStop}%
\end{thebibliography}

\end{document}